\documentclass[prl,showpacs,twocolumn,aps]{revtex4}
\usepackage{graphicx}
\newcommand{\be}{\begin{equation}}
\newcommand{\ee}{\end{equation}}
\newcommand{\bea}{\begin{eqnarray}}
\newcommand{\eea}{\end{eqnarray}}

\newcommand{\bwt}{\begin{widetext}}
\newcommand{\ewt}{\end{widetext}}
\begin{document}
\title{
How many electrons are needed to flip a local spin?
}
\author{Wonkee Kim, R.K. Teshima, and F. Marsiglio}
\affiliation{
Theoretical Physics Institute and
Department of Physics, University of Alberta, Edmonton, Alberta,
Canada, T6G~2J1}
\begin{abstract}
Considering the spin of a local magnetic atom as a quantum mechanical operator,
we illustrate the dynamics of a local spin interacting with a ballistic
electron represented by a wave packet. This approach improves the
semi-classical approximation and provides a complete quantum mechanical
understanding for spin transfer phenomena. Sending spin-polarized electrons 
towards a local magnetic atom one after another,
we estimate the minimum number of electrons needed to flip a local spin.
\end{abstract}

\date{\today}
\maketitle

Electrons generally interact with magnetic atoms through a spin-flip
interaction. For example, this is a fundamental mechanism for
spin transfer \cite{slonczewski,berger}
from spin-polarized electrons to a magnetic moment in a
ferromagnet \cite{myers,tsoi,kiselev}.
Spin transfer results in
a classical torque exerted on the magnetic moment
and, thus, enables us to control a local structure in a
ferromagnetic film.
This picture is semi-classical and
seems to work well n a practical sense \cite{bazaliy,str,mese,kim2}.
However, this view also causes some conceptual difficulties,
and cannot answer fundamental questions associated with spin transfer.
For example, it seems intuitive that the stronger
the spin-flip interaction is, the easier it is to flip a local spin
through the interaction. However, a more rigorous quantum mechanical
treatment of the problem will illustrate that this intuition is
incorrect.

The goals in this paper are following: (1) we scrutinize
the physics of spin transfer between an incoming electron
and a localized magnetic atom with spin $S$, {\it i.e.} we calculate
the expectation value of the local spin operator as a function 
of time $(t)$ and illustrate the physics in detail. (2) We provide
an estimation of how many electrons are needed to flip the local spin by
sending spin-polarized electrons one after another with an interval 
sufficiently
long that no interference between two electrons occurs.
For illustrative purposes, we consider spin values of $S=1/2,\;1$, and $3/2$
to represent the local spin.

Imagine a magnetic atom residing at the origin of the $X$ axis with
spin pointing to the negative $Z$ axis. Its spin state can be represented
by $|-S\rangle$ at $t=0$. We consider 
a normalized wave packet $\phi$ of an electron with spin along
the positive $Z$ direction
away from the origin. Then, the total
wave function is initially $\Psi=\phi |+\rangle|-S\rangle$, where
$|+\rangle$ is the spin state of an incoming electron. Since the local atom
is neutral, the Coulomb interaction will not be included.
This problem can be set up in one dimension.
The Hamiltonian we consider is
\be
H=\frac{p^{2}}{2m}-2J_{0}{\bf \sigma}\cdot{\bf S}\delta(x)
\ee
where $p^{2}/2m$ is the kinetic energy of the incoming electron,
${\bf \sigma}$ is the electron spin operator, and
$J_{0}\; (>0)$ is the coupling of the spin-flip interaction. 
We use units such that $\hbar=c=1$.
Let us introduce $\lambda/a=2mJ_{0}$, 
where $a$ is a typical length scale
in the problem and will be set to unity.
The initial wave packet is normalized and given by
$
\phi(x)=(2\pi\alpha)^{-1/4}e^{ik_{0}(x+x_{0})}
e^{-(x+x_{0})^{2}/4\alpha}
$. 
Such a wave packet describes an "electron" with mean position $-x_{0}$
and mean momentum $k_{0}$.
The uncertainties associated with the packet are 
$\Delta x_{0}=\sqrt{2\alpha}$ and
$\Delta k_{0}=1/\sqrt{2\alpha}$.
In order to calculate the expectation value of
the Z component of the local spin 
$\langle S_{z}(t)\rangle=\langle\Psi(x,t)|S_{z}|\Psi(x,t)\rangle$, 
where $\langle\cdots\rangle$ means an integration over $x$,
we need to know
the time evolution of the total wave function $\Psi(x,t)$.
Since the initial state $\Psi(x,0)$ is not an eigenstate of
the Hamiltonian because of the spin-flip coupling,
$\Psi(x,0)$ should be decomposed into two channels
depending on the total spin $j_{\pm}=S\pm 1/2$.
 
The Schr{\"o}dinger equation to solve is
\be
\partial^{2}_{x}\Psi(x)+2mE\Psi(x)=
\Biggl\{\begin{array}{c}
-S\lambda\delta(x)\Psi(x)\hskip 0.5cm \mbox{for} \hskip 0.5cm j_{+}\\
(S+1)\lambda\delta(x)\Psi(x)\hskip 0.5cm \mbox{for}\hskip 0.5cm j_{-}
\end{array}
\label{Schrodinger}
\ee
A wave function with $j_{+}$ sees a potential well while
a wave function with $j_{-}$ feels a potential barrier.
The time evolution of each channel is different because
the eigenstates of each equation are different, and the overlap of the two
wave functions determines the dynamics of the local spin as
we will soon show.
The time evolution of the wave function is a
combination of the time evolutions of the two channels of $j_{\pm}$:
\bwt
\be
\Psi(x,t)=\phi_{+}(x,t)\frac{1}{\sqrt{2S+1}}|j_{+},-S+1/2\rangle
+\phi_{-}(x,t)\sqrt{\frac{2S}{2S+1}}|j_{-},-S+1/2\rangle
\ee
\ewt
where $\phi_{\pm}(x,t)$ is the time evolution of the wave packet
in the presence of the well/barrier, and $\phi_{\pm}(x,0)=\phi(x)$.
The $Z$ component
of the total spin is conserved to be $-S+1/2$.
The expectation value of the Z component of the local spin 
as a function of time
is calculated to be
\be
\langle S_{z}(t)\rangle=-\frac{S}{(2S+1)^{2}}\left\{
4S^{2}+4S+4\mbox{Re}\left[\langle\phi_{+}|\phi_{-}\rangle\right]-3
\right\}\;.
\label{S_1}
\ee
Using the Schwartz inequality: $|\langle\phi_{+}|\phi_{-}\rangle|^{2}\le 1$,
we find a constraint of $\langle S_{z}(t)\rangle$
as follows:
$-S \le \langle S_{z}(t)\rangle \le -S\left[4S^{2}+4S-7\right]/(2S+1)^{2}$.
This constraint is more restrictive than the obvious one that requires
a maximum change in the spin to be unity since the spin of the
incoming electron is $1/2$.
Note that there always exists
a probability that the state of the local spin remains 
unchanged even after the interaction; therefore, actual values of
$\langle S_{z}(t)\rangle $ should be evaluated quantum mechanically.
We can also determine the local spin state from the total wave function
$\Psi(t)$. The spin state is written as 
$a_{1}|-S\rangle+b_{1}|-S+1\rangle$, where
$|a_{1}|^{2}=\left\{1+4S^{2}+4S\mbox{Re}
\left[\langle\phi_{+}|\phi_{-}\rangle\right]\right\}/(2S+1)^{2}$ and
$|b_{1}|^{2}=4S\left\{1-\mbox{Re}
\left[\langle\phi_{+}|\phi_{-}\rangle\right]\right\}/(2S+1)^{2}$.
It is straightforward to show that $\langle S_{z}(t)\rangle
=-S+|b_{1}|^{2}$. Without loss of generality we can assume that
$a_{1}$ and $b_{1}$ are real.

As mentioned for $\langle S_z(t)\rangle$, we need
to solve Eq.~(\ref{Schrodinger}) because
the time evolution of the wave packet
$\phi_{\pm}$ will be governed by the solutions of this equation.
Let us consider a Hamiltonian 
$H_{\pm}=p^{2}/2m+(\lambda_{\pm}/2m)\delta(x)$,
where $\lambda_{+}=-S\lambda$ for the $j_{+}$ channel while
$\lambda_{-}=(S+1)\lambda$ for the other channel.
This Hamiltonian could be treated as a scattering problem
for $E>0$
to obtain an asymptotic solution which is plane wave-like.
Note that the asymptotic solution does not describe the detailed
dynamics of the local spin. However, it implies
important physics associated with the problem. The asymptotic
solution is $\eta_{k}(x)=\left[e^{ikx}+R_{\pm}e^{-ikx}\right]\Theta(-x)
+T_{\pm}e^{ikx}\Theta(x)$, where $\Theta(x)$ is a step function. 
The momentum is well-defined
while the position cannot be; namely, 
$\Delta k=0$ but $\Delta x=\infty$. The transmittance and the reflectance
are determined by the boundary condition at $x=0$, and they are
$|T_{\pm}|^{2}=4/\left[4+(\lambda_{\pm}/k)^{2}\right]$ and
$|R_{\pm}|^{2}=(\lambda_{\pm}/k)^{2}/\left[4+(\lambda_{\pm}/k)^{2}\right]$.
It is worth mentioning that $|T_{\pm}|^{2}$ and $|R_{\pm}|^{2}$
depend only on $(\lambda_{\pm}/k)^{2}$; furthermore, for large
$\lambda_{\pm}/k$ both channels give unit reflectance and almost zero
transmission.
As we showed in Eq.~(\ref{S_1}), the dynamics of the local spin
depends on the overlap of the two wave functions $\phi_{+}(x,t)$
and $\phi_{-}(x,t)$. The above exercise indicates that for a large
$\lambda$, $\phi_{\pm}$ would not differ significantly from each other.
In this instance, 
$\mbox{Re}\left[\langle\phi_{+}|\phi_{-}\rangle\right]\approx1$;
consequently, $\langle S_z \rangle \approx -S$. 
In other words, if the spin-flip coupling is very large,
it becomes more difficult to flip a local spin. 
This seems to oppose the semi-classical understanding but 
we will show that this is the case.
We should mention that the $\lambda_{\pm}/k$ scaling in 
$|T_{\pm}|^{2}$ and $|R_{\pm}|^{2}$ takes place
because of zero uncertainty in the momentum $(\Delta k=0)$ for a plane wave.
Rigorously speaking, however, 
one cannot expect such a perfect scaling when a wave packet with
$\Delta k\ne0$ is used instead
of a plane wave.

We obtain the eigenstates
and the corresponding eigenvalues of $H_{\pm}$  
introducing a box of the length $2L$ $(-L\le x \le L)$
with a periodic boundary condition: $\psi(-L)=\psi(L)$
and $\partial_{x}\psi(-L)=\partial_{x}\psi(L)$, where
$\psi(x)$ is an eigenstate of $H_{\pm}$ with an eigenvalue 
$E_{\pm}$. Since the potential is symmetric about $x=0$,
the eigenstates are either even or odd; $\psi_{e,k_{\pm}}(x)$ or
$\psi_{o,p_{\pm}}(x)$. 
We found \cite{we_did_it} for the even solution
\be
\psi_{e,k_{\pm}}(x)=\frac{1}{\sqrt{N_{k_{\pm}}}}
\left[\cos(k_{\pm}x)+\frac{\lambda_{\pm}}{2k_{\pm}}\sin(k_{\pm}|x|)\right]
\ee
where $N_{k_{\pm}}=L\left[1+(\lambda_{\pm}/2k_{\pm})^{2}\right]+
\lambda_{\pm}/2k^{2}_{\pm}$,
and for the odd solutions
$\psi_{o,p_{\pm}}(x)=\sin(p_{\pm}x)/\sqrt{L}$.
The corresponding eigenvalues are $E_{k_{\pm}}=k^{2}_{\pm}/2m$, where
$k_{\pm}$ is a solution of $\tan(k_{\pm}L)=\lambda_{\pm}/2k_{\pm}$,
while $E_{p_{\pm}}=p^{2}_{\pm}/2m$ with $p_{\pm}=n_{\pm}\pi/L$ 
($n_{\pm}$ is an integer).
Since the potentials do not affect the odd solutions, the dynamics of
the local spin is determined only by the even solutions.
For $\lambda_{+}\;(<0)$, the number of the even states decreases by one and
a single bound state occurs for $E<0$ to keep the total number of 
eigenstates unchanged.
As long as we initially put a wave packet far away from $x=0$, 
the bound state does not
participate in the time evolution of the wave packet \cite{qm}. 
Now the time evolution of $\phi_{\pm}(x,t)$ for both channels can be written
as
\be
\phi_{\pm}(x,t)=
\sum_{k_{\pm}}e^{-i E_{k_{\pm}}t}C_{k_{\pm}}\psi_{e,k_{\pm}}(x)
+\sum_{p_{\pm}}e^{-i E_{p_{\pm}}t}C_{p_{\pm}}\psi_{o,p_{\pm}}(x)
\ee
where $C_{k_{\pm}}=\langle \psi_{e,k_{\pm}}|\phi\rangle$ and
$C_{p_{\pm}}=\langle \psi_{o,p_{\pm}}|\phi\rangle$. Using these expressions
we obtain the overlap between $\phi_{+}$ and $\phi_{-}$ as follows:
\bea
\langle\phi_{+}|\phi_{-}\rangle&=&
(2S+1)\lambda\sum_{k_{\pm}}\frac{e^{i(E_{k_{+}}-E_{k_{-}})t}}
{k^{2}_{-}-k^{2}_{+}}
\frac{\langle\phi|\psi_{e,k_{+}}\rangle}{\sqrt{N_{k_{+}}}}
\frac{\langle\psi_{e,k_{-}}|\phi\rangle}{\sqrt{N_{k_{-}}}}
\nonumber\\
&+&\frac{1}{2}
\left[1-e^{-(k_{0}/\Delta k_{0})^2-(x_{0}/\Delta x_{0})^2}\right]\;,
\label{overlap}
\eea
where $\Delta k_{0}$ and $\Delta x_{0}$ are the uncertainties
associated with the initial wave packet.
Note that the time dependence of $\langle\phi_{+}|\phi_{-}\rangle$
is determined only by the even solutions
while the odd solutions contribute to 
a constant, which would be close to $1/2$ 
if the momentum or the position is well defined and finite initially.
Since we consider 
a wave packet far away from the origin with a finite mean position at $t=0$,
the second term in Eq.~(\ref{overlap}) is approximately $1/2$ to high 
precision.

Fig.~1 illustrates the time evolution of the wave packet for the $j_{-}$
channel away from a local spin of $S=1/2$. 
We introduce a dimensionless time $\tau=t/2ma^{2}$. Initially
the wave packet is located at $x=-x_{0}$ $(x_{0}=100)$ 
with a mean momentum $k_{0}=1$.
We choose $\alpha$ to be $10$ and $L$ to be $10^{3}\sim 10^{4}$.
As long as $L\gg x_{0}$, $L$ is not important.
The wave packet does not interact with the potential with 
$\lambda_{-}=(S+1)\lambda=3.45$ until $\tau\simeq30$. At $\tau=50$,
the potential strongly scatters
the wave packet and a part of the packet is transmitted to 
the other side of the potential. After $\tau=150$, the reflected and
transmitted parts of the packets freely move away from the potential
in the opposite direction as indicated by the arrows. 
The time evolution for $j_{+}$ is similar and not shown in this figure.
In the insert,
we show $\langle S_{z}(\tau)\rangle$ as a function of $\tau$.
It increases from $\langle S_{z}(0)\rangle=-0.5$ to 
$\langle S_{z}(150)\rangle\simeq-0.01$, and then becomes saturated.
The saturation is expected because after $\tau=150$ the wave packet does not
interact with the potential any more. 
In fact, the mean momentum determines
when $\langle S_{z}\rangle$ becomes saturated.
We define the saturation value of $\langle S_{z}\rangle$ as $S_{1}$, which 
depends on $k_{0}$ and $\lambda$. It is this value that we use to estimate
how many spin-polarized electrons are needed to flip a local spin
as we will show later.

It is possible to control the mean momentum reasonably well while 
the spin-flip coupling is uncontrollable and even not well known 
experimentally.
In this sense, it does not seem possible to pinpoint how many electrons
are needed to flip a specific local spin. Nevertheless,
it is theoretically feasible to estimate the minimum number 
of electrons to flip a local spin by calculating the maximum value of
$S_{1}$ denoted as $S^{max}_{1}$. To this end, 
we evaluate $S_{1}$ as a function of $k_{0}$ 
and $\lambda$ because a scaling relation is not expected to hold
for the wave packet with the momentum uncertainty $(1/\sqrt{2\alpha})$.
We choose the same value of $\alpha$ as before.
Fig.~2 is a contour plot of $S_{1}$ in the $(\lambda,\; k_{0})$ plane.
Interestingly, we found a $\lambda/k_{0}$ scaling
holds fairly well in this case as well.
Moreover, $S^{max}_{1}$ of a local spin $S=1/2$ is approximately zero 
for $\lambda\simeq2.3k_{0}$; for example, $\lambda=6.9$ with $k_{0}=3$. 
We obtained the same scaling
behavior for other values of $\alpha$.
Note that this value of $S^{max}_{1}$
is smaller than the mathematical upper bound based on the
Schwartz inequality, which is a half for $S=1/2$.
Looking in particular at the lower part of Fig.~2, we know that 
for a given $k_{0}$, $S_{1}$ decreases with increasing $\lambda$. 
This clearly indicates that
if the coupling is too large, it becomes more and more difficult
to flip the local spin, which is consistent with
our early analysis. We plot $S_{1}$ for the local spin $S=1$ and $S=3/2$
in Fig.~3 as a function of $\lambda/k_{0}$ with $\alpha=10$. 
There is also a similar
scaling behavior in $S_{1}$ while different values of $\lambda/k_{0}$
give $S^{max}_{1}$. For $S=1$, the maximum value
is $S^{max}_{1}\simeq-0.556\; (<-1/9)$ along $\lambda\simeq1.4k_{0}$, and 
$S^{max}_{1}\simeq-1.126\; (< -3/4)$ along $\lambda\simeq k_{0}$
for $S=3/2$, where $-1/9$ and $-3/4$ are the mathematical upper bounds
for $S=1$ and $S=3/2$, respectively.

Now consider the spin-polarized electrons sent towards the local spin
one after another over an interval $\tau_{0}$ which is sufficiently long
so as to prevent any interference
between two electrons. That is, we wait long enough before sending the
second electron so that the first has cleared away from the region of
interest, and left the local spin in a state
$a_{1}|-S\rangle+
b_{1}|-S+1\rangle$, where $a_{1}$ and $b_{1}$ have acquired saturated
values after a time $\tau_0 \approx 200$ (see Fig.~1). Note that this time
can be shorter if we increase the mean momentum, for example.
At $\tau=\tau_{0}$, the total wave function
is $\phi|+\rangle
\Bigl[a_{1}(\tau_{0})|-S\rangle+b_{1}(\tau_{0})|-S+1\rangle\Bigr]$.
Since $|+\rangle|-S\rangle$ and $|+\rangle|-S+1\rangle$ are not
eigenstates of the Hamiltonian in general, we need to decompose
each state into two channels for $j_{\pm}$ with appropriate Clebsch-Gordan
coefficients \cite{schiff}. Later the spin state of the local spin will be
$a_{2}|-S\rangle+b_{2}|-S+1\rangle+c_{2}|-S+2\rangle$.
The coefficients $a_{2}$, $b_{2}$, and $c_{2}$ can be determined from
the total wave function, and they turn out to be functions of 
$a_{1}$ and $b_{1}$.
Repeating this procedure we can calculate the expectation value
of the local spin after sending the $n$-th electron. 

Using $S=1/2$, $1$ and $3/2$ we illustrate the procedure and
estimate the minimum number of electrons to flip a local spin.
The expansion for an arbitrary spin can be done systematically.  
For a local spin $S=1/2$, $j_{+}=1$ and $j_{-}=0$. 
After the first electron interacts with
the local spin, the total wave function is
$\Psi=\frac{1}{\sqrt{2}}\phi_{1}|10\rangle+
\frac{1}{\sqrt{2}}\phi_{0}|00\rangle$. The spin state of the atom
is $a_{1}|-1/2\rangle+b_{1}|1/2\rangle$, where $a^{2}_{1}=1/2-S_{1}$ and
$b^{2}_{1}=1/2+S_{1}$, and $\langle S_{z} \rangle=S_{1}$. 
When we send the second
electron, $\Psi=\frac{1}{\sqrt{2}}a_{1}\phi_{1}|10\rangle+
\frac{1}{\sqrt{2}}a_{1}\phi_{0}|00\rangle+b_{1}\phi_{1}|11\rangle$.
The local state becomes $a_{2}|-1/2\rangle+b_{2}|1/2\rangle$, where
$a^{2}_{2}=a^{4}_{1}$ and $b^{2}_{2}=\left(a^{2}_{1}+1\right)b^{2}_{1}$,
and $S_{2} = 1/4+S_{1}-S^{2}_{1}$, where $S_{n}$ is $\langle S_{z} \rangle$
after the $n$-th electron. 
The third electron gives
$S_{3}=1/4+\left(S_{1}+S_{2}\right)/2-S_{1}S_{2}$, and the $n$-th electron
leaves $S_{n}=1/4+\left(S_{1}+S_{n-1}\right)/2-S_{1}S_{n-1}$. We can
express $S_{n}$ in terms of $S_{1}$ as follows:
\be
S_{n}=\frac{1}{2}-\left(\frac{1}{2}-S_{1}\right)^{n}\;.
\ee
Using a similar procedure we obtain $S_{n}$ for $S=1$ and $S=3/2$
\bea
S_{n}&=&1-n\left(-S_{1}\right)^{n-1}+(n-2)\left(-S_{1}\right)^{n}
\hskip 0.3cm\mbox{for}\hskip 0.3cm S=1
\nonumber\\
S_{n}&=&\frac{3}{2}-\left(-S_{1}-1/2\right)^{n-1}\Bigl[
(3+4n)S_{1}+3(1+4n)/2\Bigr]
\nonumber\\
&-&6\left(-4S_{1}/3-1\right)^{n}
\hskip 0.3cm\mbox{for}\hskip 0.3cm S=3/2
\eea
Note that $S_{0}=-S$, which means that
initially the local spin state is $|-S\rangle$ 
while $S_{n}\rightarrow S$ as $n$ increases
for a given $S_{1}$; in other words, the local spin becomes 
flipped. Mathematically speaking, $S_{n}= S$ only when $n=\infty$.
This is because there always exists a quantum mechanical 
possibility that the spin state remains unchanged even after
the spin-flip interaction. 
Nevertheless, when $S_{n}$ becomes sufficiently
close to $S$, we can claim that the local spin has flipped.
When $S_{1}=S^{max}_{1}$,
we can approximate
$S^{max}_{n}\simeq S-2S\;e^{-\beta_{S}n}$,
where $\beta_{1/2}\simeq0.7$, $\beta_{1}\simeq0.35$, and 
$\beta_{3/2}\simeq0.23$.
Let us define the minimum number $N_{S}$ 
to be the number
which satisfies, for example,
$2S\;e^{-\beta_{S}N_{S}}=10^{-5}$.
Then we evaluate $N_{S}=\left[5\ln(10)+\ln(2S)\right]/\beta_{S}$.
Our estimation shows that the minimum number of the spin-polarized 
electrons to flip a local spin of $1/2$ is about $16$; 
namely, $N_{1/2}\simeq16$.
For $S=1$, $N_{1}\simeq35$ while 
$N_{3/2}\simeq55$ for the local spin of $3/2$.
Therefore, less electrons are needed to flip a smaller spin
as one might expect.

In summary, using straightforward quantum mechanics, 
we have studied the time evolution of
the spin of a local magnetic atom
under a spin-flip interaction with an incoming electron. This treatment goes
beyond the semi-classical approximation, which considers
the local moment as a classical vector. The expectation value of the spin operator
has been evaluated using the wave function of the electron, which
is the solution of the time dependent Schr{\"o}dinger equation.
Sending spin-polarized electrons
towards a local magnetic atom one after another,
we also provide an estimate of how many electrons are needed 
to flip a local spin.
For an experimental realization of our estimate,
we suggest a setup where
a magnetic atom is fixed at the hub of a wheel, while spin-polarized electrons
are sent towards the atom along orthogonal spokes in the wheel.

One of us (W.K.) thanks H.K. Lee and M. Revzen for discussions.
This work was supported in part by the Natural Sciences and Engineering
Research Council of Canada (NSERC), by ICORE (Alberta), and by the
Canadian Institute for Advanced Research (CIAR).
\bibliographystyle{prl}

\begin{thebibliography}{1}
\bibitem{we_did_it} As far as we know, these eigenstates and the
corresponding eigenvalues have never
been presented in literature.
\bibitem{qm} Mathematically speaking, a wave function $\phi(x,t)$ is expanded
in terms of eigenstates $u_{n}(x)$ with eigenvalues $E_{n}$;
$\phi(x,t)=\sum_{n}C_{n}(t)u_{n}(x)$, where $C_{n}(t)=C_{n}(0)e^{-iE_{n}t}$.
Therefore, if $C_{n'}(0)=0$, then $C_{n'}(t)=0$.
\end{thebibliography}

\begin{figure}[tp]
\begin{center}
\includegraphics[height=3.5in,width=3.5in]{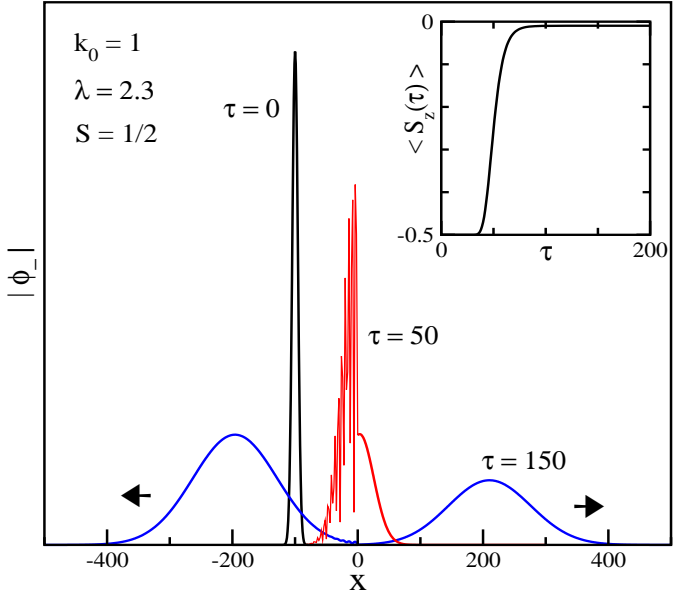}
\caption{(Color online)
The time evolution of the wave packet for the $j_{-}$ channel.
The other channel shows similar behavior and is not plotted. The local spin is
$1/2$. The inset describes dynamics of the local spin.
After $\tau=150$, $\langle S_{z}\rangle$ becomes saturated. }
\end{center}
\end{figure}

\begin{figure}[tp]
\begin{center}
\includegraphics[height=3.5in,width=3.5in]{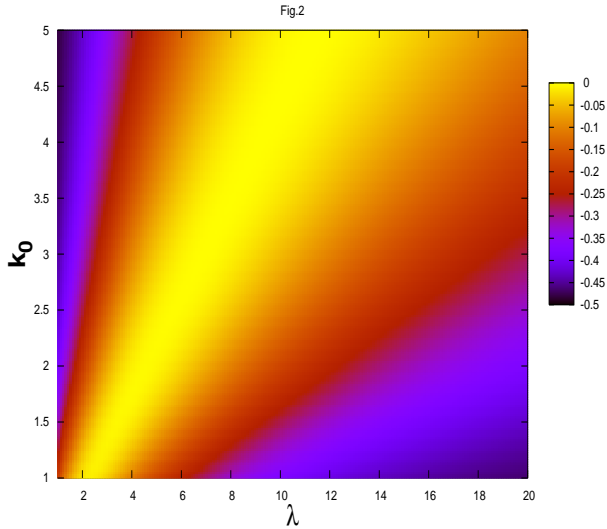}
\caption{(Color online)
Contour plot of $S_{1}$ for $S=1/2$
in the $(\lambda,\; k_{0})$ plane.
Note a $(\lambda/k_{0})$ scaling behavior with a 
maximum value of $S_1$ along $\lambda=2.3k_{0}$. }
\end{center}
\end{figure}

\begin{figure}[tp]
\begin{center}
\includegraphics[height=3.5in,width=3.5in]{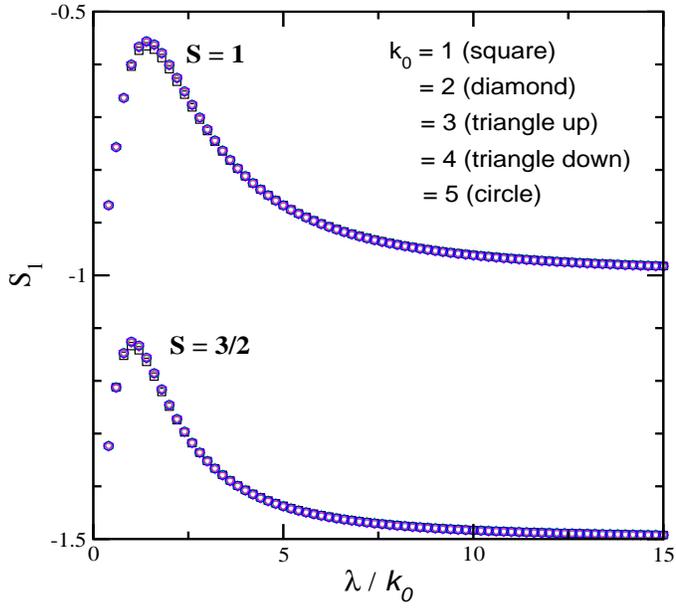}
\caption{(Color online)
$S_{1}$ as a function of $\lambda/k_{0}$
for $S=1$ and $S=3/2$, for various values of $k_0$. The maximum value of $S_{1}$ is about $-0.556$
and $-1.126$ for $S=1$ and $S=3/2$, respectively. }
\end{center}
\end{figure}
\end{document}